\documentclass{ws-p10x7}
\newcommand{\qprimesq}{\mbox{$Q'^2~$}}
\newcommand{\xprime}{\mbox{$x'~$}}

\newcommand{\GeVsqx}{\mbox{${\rm GeV}^2$}}

\newcommand{\pb}{\mbox{${\rm ~pb}~$}}

\begin{document}
\title{Search for QCD-instantons at HERA}
\author{E.A.~De Wolf}

\address{CERN, European Organisation 
for Nuclear Research,
CH-1211 Geneva 23, Switzerland and\\
Universitaire Instelling Antwerpen,  Universiteitsplein 1, B-2610 Antwerpen 
\\E-mail: eddi.de.wolf@cern.ch\\[.3cm]
 {\rm FOR THE H1 COLLABORATION}}
\twocolumn[\maketitle\abstract{
Signals of QCD instanton induced processes are searched for in deep-inelastic 
$ep$ scattering at HERA in a kinematic region defined  by the Bjorken scaling variables
$x>10^{-3}$, $0.1<y<0.6$ and polar angle of the scattered positron $\theta_{el}>156^\circ$. 
Upper limits  are derived from  the expected instanton-induced
final state properties based on the QCDINS Monte Carlo model.}]

\section{Instantons in DIS}
After initial work by Balitsky and Braun\cite{bali}, theoretical 
and pheno\-meno\-logic\-al  studies of the r\^ole 
of instantons in deep inelastic scattering (DIS) at HERA have been vigorously 
pursued\footnote{See papers 250--252, 254, 255 submitted to this Conference.} by Ringwald, 
Schrempp and collaborators\cite{sch}.

\begin{figure}[t]
\epsfxsize6.7cm  
\figurebox{6.7cm}{5cm}{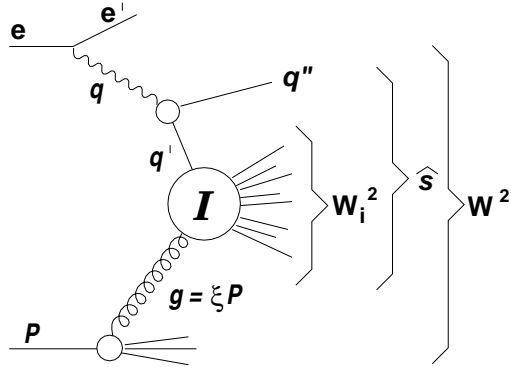}
   \label{fig:diagram}
   \caption{\small Dominant graph for the instanton-induced contribution to deep-inelastic
              $ep$ scattering with the relevant kinematic variables as indicated: $Q'^2 = -q'^2 =
              -(q-q')^2$, $x'=Q'^2/2(g\cdot q')$, $W_i^2 = (q'+g)^2=Q'^2
              (1-x')/x'$.}
\end{figure}

Instanton induced processes (I-events) in DIS arise predominantly from  photon gluon
fusion in an instanton background (see Fig.~\ref{fig:diagram}) via the reaction
$$
\gamma^* + g \to \sum_{n_{\rm flavours}} (\bar{q}_R + q_R) + \, n_g \, g,
$$
where $q_R$ ($\bar{q}_R$) denotes right handed quarks\footnote{%
Right handed quarks are produced in instanton processes,
left-handed ones  in anti-instanton processes.}
and $g$ gluons.
The cross section is calculated\cite{sch2000} to be in the range 10--100~pb 
for  $0.1 < y < 0.9$ and $x > 10^{-3}$, which is sizable 
but three orders of magnitude smaller than that of ``normal'' DIS events.

The final states in I-events are characterized by:
a current-quark jet ($q^{\prime\prime}$, Fig.~\ref{fig:diagram}), 
a partonic final state from I-decay
consisting  of $2 n_f - 1$  right-handed quarks and anti-quarks.
In every I-induced event, one quark anti-quark pair
of all $n_f(=3)$ flavours is simultaneously
produced.
In addition, on average $\langle n_g \rangle ^{(I)} \sim {\cal O}(1/\alpha_s)  \sim  3$
gluons are isotropically emitted in the I-process.
I-events are thus expected to show a
pseudo-rapidity
  ($\eta$)
region (with a width of about 1.1 units)
densely populated with particles of high transverse momentum and
uniformly distributed in azimuth.
This, together with the high density of partons emitted in the I-process leads
to a high particle multiplicity and large transverse energy.
To simulate QCD-instanton induced scattering processes in DIS and their characteristic
final states, the QCDINS  Monte Carlo generator
was developed.\cite{qcdins}
 It is based on instanton perturbation theory and imbedded in HERWIG.\cite{herwig}

\section{H1 results}
The preliminary H1 results presented  here use   data taken 
in 1997 with the H1 detector, corresponding to
an integrated  luminosity  of  $ {\cal L} =15.78~\rm pb^{-1}$.
The analysis is performed in the DIS kinematic region
  $0.1 < y_{el} < 0.6$,   $x_{el} > 10^{-3}$
and $\theta_{el} > 156^\circ$. 
 The  total DIS sample comprises $\sim 280$K events.

The search strategies aim to enrich a data sample in $I$-induced events 
using cuts on selected observables while optimizing the
separation power, defined as the ratio of the detection efficiencies, for
$I$-induced and DIS-events.

The following observables\footnote{All observables, except sphericity, are calculated 
in the hadronic CMS ($\vec q + \vec P = \vec 0$).} have 
been used:
(1) $Et_{jet}$, the jet with highest $E_T$ (cone algorithm 
with radius $R=0.5$). This jet is associated with the current
quark (q") in Fig.~\ref{fig:diagram}.   
(2)  The virtuality of the quark entering the I-process 
${Q'}^2 = -(q-q")^2$    
 where the photon (4-momentum $q$) is reconstructed from  the scattered
electron.
(3)  The number of charged particles $n_B$ in the so-called 
``instanton band''.\footnote{%
Particles belonging to the jet $q"$ are removed from the final state
and
the $E_T$-weighted mean pseudorapidity $\bar{\eta}$
is recalculated with the remaining ones.
The instanton band is  defined as ~$\bar{\eta} \pm 1.1$.}          
(4)  The sphericity SPH  calculated in the rest system of the
particles outside the current jet. 
(5) $Et_b$  the total transverse energy in the
instanton band calculated  
as the scalar sum of the transverse energies
and (6) $\Delta_b$,\footnote{ $\Delta_b$ is defined as 
$
\Delta_b = \frac{E_{in} - E_{out}}{E_{in}}
$
where $E_{in}$ ($E_{out}$) is the maximal (minimal) value of 
the sum of the projections  on all possible axis $\vec{i}$
of all energy depositions in the band 
(i.e. $E_{in}= max \sum_n |\vec{p_n} \vec{i} |$). 
For isotropic events (jet-like events) $\Delta_b$ is expected to be small (large).} 
a quantity measuring the $E_T$ weighted $\Phi$ event isotropy. 

Among many studied, three scenarios  
are chosen based on the following criteria:
(A) The highest instanton efficiency ($\epsilon_{ins}$ $\approx 30 \%$),  
(B) high $\epsilon_{ins}$ with reasonable background reduction and  
(C) highest background reduction ($\epsilon_{dis}$ $\approx 0.13-0.16 \%$)
      with   $\epsilon_{DIS}$ $\approx 10 \%$.

\begin{table*}[t]
\caption{Measured numbers of events and expected background
 for  3 cut scenario's. The errors are systematics dominated.}
\vspace{0.2cm}
\begin{center}
\footnotesize
\begin{tabular}{|c|c|c|c|c|c|}
\hline
\multicolumn{2}{|c|} {\raisebox{0pt}[12pt][6pt]{ (A) DATA: $3000$ }} &
\multicolumn{2}{|c|}{\raisebox{0pt}[12pt][6pt] {(B) DATA: $1332$} } &
\multicolumn{2}{|c|}{\raisebox{0pt}[12pt][6pt] {(C) DATA: $ 549$} }\\
\hline
\raisebox{0pt}[12pt][6pt] {CDM} & {MEPS}  &    
 {CDM} & {MEPS}  & 
 {CDM} & {MEPS}  \\ 
\hline
 \raisebox{0pt}[12pt][6pt] {$2469^{+242}_{-238}$} & {$2572^{+237}_{-222}$} &
  {$1005^{+82}_{-70}$} & {$1084^{+75}_{-46}$} &
  {$ 363^{+22}_{-26}$} & {$435^{+36}_{-22}$} \\
\hline
\end{tabular}
\end{center}
\label{table:result}
\end{table*}

\begin{figure}[t]
\epsfxsize\columnwidth
\figurebox{\columnwidth}{}{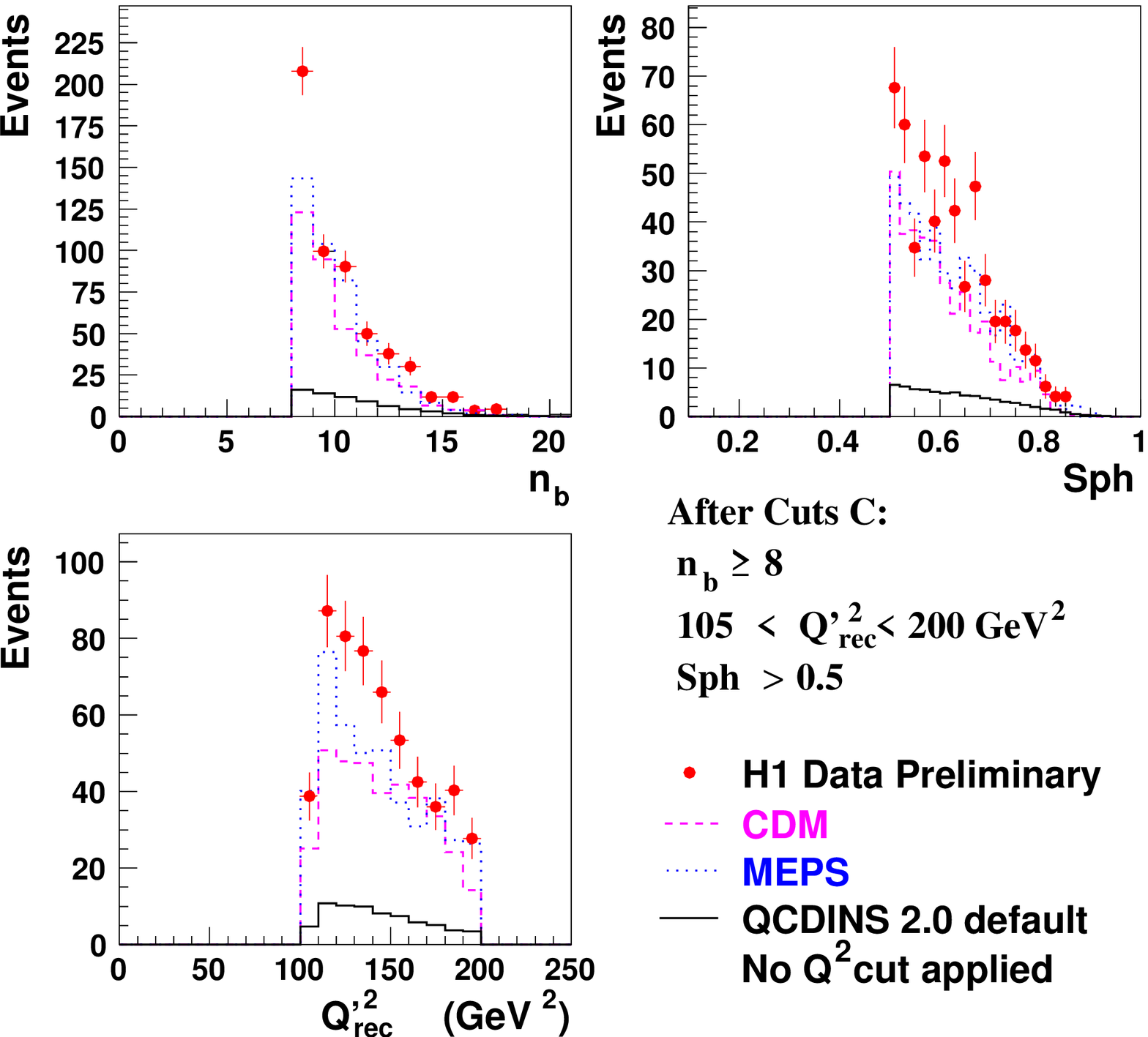}
\epsfxsize\columnwidth
\figurebox{\columnwidth}{}{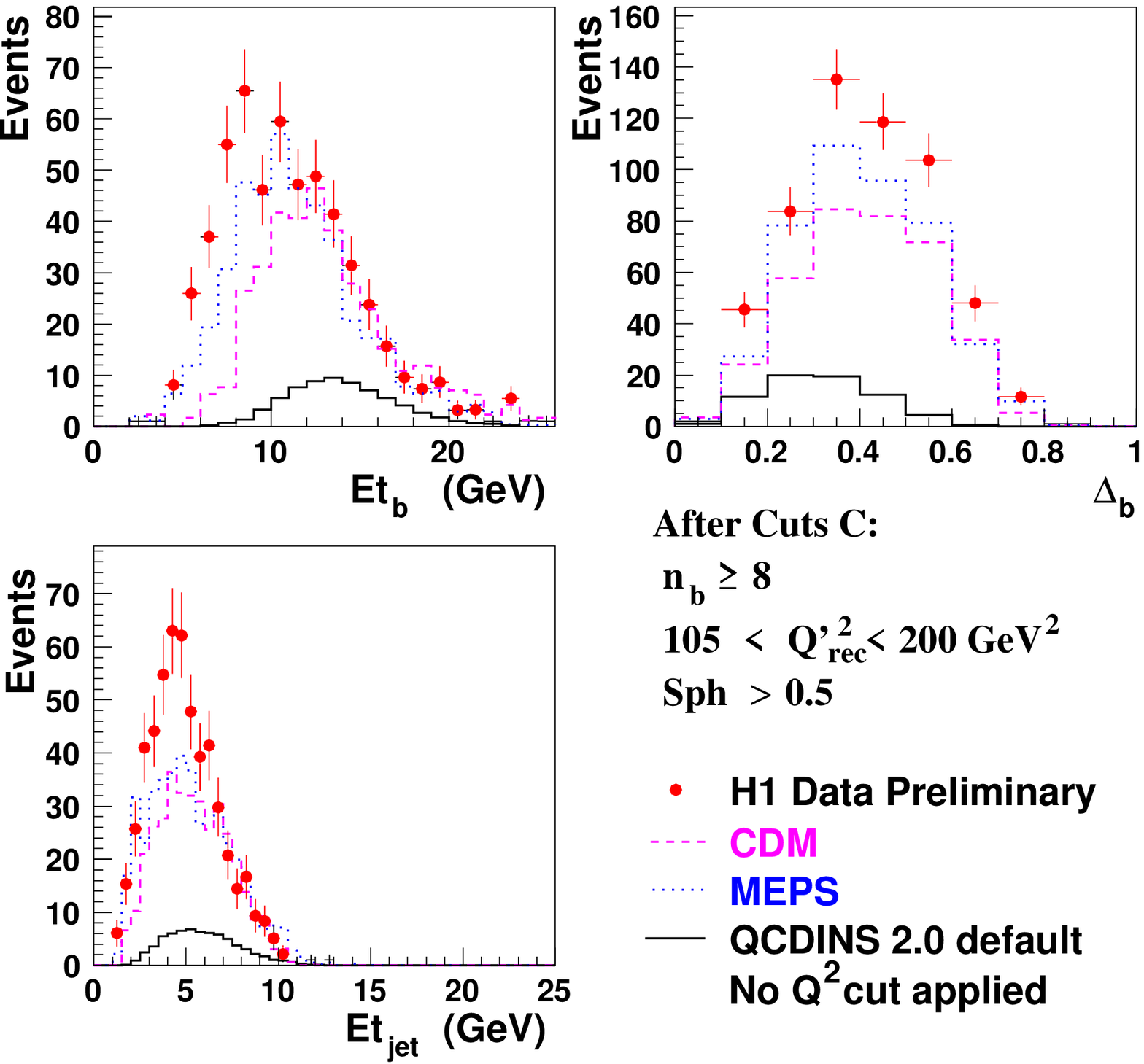}
\caption{The distributions of kinematic variables in  cut-scenario C, compared with
the Color Dipole Model (CDM)\protect\cite{ariadne}, MEPS\protect\cite{rapgap} 
and QCDINS.  \label{fig:aftercutC}}
\end{figure}  

Table~\ref{table:result} summarizes the number of events in data and expected in the standard
DIS Monte Carlo simulations after applying cuts A)--C).
Distributions after cuts~C  are shown in Fig.~\ref{fig:aftercutC}. 
An excess of events over DIS Monte Carlos is observed. 
However, the size of this
signal is comparable with the difference  between CDM and MEPS. 
Also, the excess in transverse energy $Et_b$ in the instanton band 
differs from the QCDINS expectation. Nevertheless, 
given the uncertainties in the I-event cross section
calculation and the modelling of the hadronic final state, an I-signal of the form predicted by 
QCDINS cannot be excluded at this stage of the analysis.

\begin{figure}[t]
\epsfxsize\columnwidth
\figurebox{\columnwidth}{10cm}{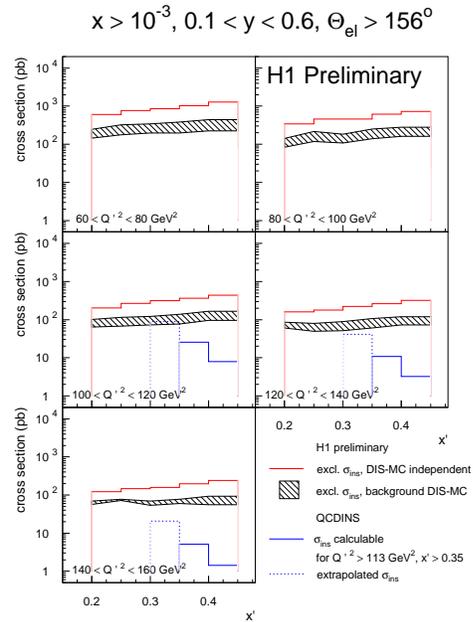}
\caption{Upper limit on the cross section for instanton induced
events as modelled by  QCDINS  as a function
of \xprime in bins of \qprimesq. 
Regions above the curves are excluded.
Also shown is the instanton cross section predicted in the
fiducial region $\xprime > 0.35$ and $\qprimesq > 113$ \GeVsqx. 
\label{fig:slim}}
\end{figure}

%
%
H1 has derived cross section upper limits ($95\%$ CL) by  comparing
the QCDINS predicted cross section, data and CDM/MEPS expectations.
Results are shown in Fig.~\ref{fig:slim}. Regions above the curves
are excluded\footnote{For a detailed description of the methods used, the reader should consult
the original H1 paper~\cite{h1:paper}.}.
Instanton cross sections between $100$ and $1000$~\pb are excluded. 

%

\section*{Acknowledgements}
I am  grateful to  T.~Carli, S.~Mikocki, A.~Ringwald and F.~Schrempp
for  help and valuable comments.

\small
\bibliography{inst}

\end{document}